\documentclass[
]{ceurart}

\usepackage{listings}
\usepackage{url}
\usepackage{microtype}

\begin{document}

\copyrightyear{2026}
\copyrightclause{Copyright for this paper by its authors. Use permitted under Creative Commons License Attribution 4.0
International (CC BY 4.0).}
\conference{TETHICS 2026: 9th Conference on Technology Ethics, November 11--12, 2026, Lahti, Finland}

\title{From Smarter to Hungrier: the Role of Energy Efficiency in Software-defined Vehicles}

\author{Ella Peltonen}[%
orcid= 0000-0002-3374-671X,
email=ella.peltonen@oulu.fi
]
\address{University of Oulu, Oulu, Finland}

%\cortext[1]{Corresponding author.}

\begin{abstract}
Automated features and advanced driving support systems have enabled a range of life-saving and comfort applications in modern personal vehicles. Development from mostly mechanical entities towards truly software-defined devices also means that added sensor complexity, processing power, and networking capabilities provide a platform for running even more complex algorithms and applications. As vehicles become smarter, they require even more energy to enable advanced machine learning- and artificial intelligence-based functionalities. This paper discusses the implications for sustainability and the threats to environmental ethics posed by the development of software-defined vehicles. We discuss the balance between introducing more resource-hungry software and hardware components and their effect on society and sustainability. We underscore the need for novel tools to measure, manage, and optimise energy consumption and other sustainability indicators, with design guidance and actionable recommendations. 
\end{abstract}

\begin{keywords}
  Sustainability \sep
  Software-defined vehicles \sep
  Energy-efficiency
\end{keywords}

\maketitle

\section{Introduction}

Regarding to the European Environment Agency, transportation is responsible for a quarter of carbon emissions in the European Union, out of which almost 72\% comes from road transportation \footnote{\url{https://www.europarl.europa.eu/topics/en/article/20190313STO31218/co2-emissions-from-cars-facts-and-figures-infographics}}. Electrification is considered the key driver of making transportation more sustainable, but the change towards fully electric traffic is still slow. At the same time, the increased safety demands and the rise of the software-defined vehicle (SDV) paradigm have made vehicles dependent on computational capacity, networking capabilities, and software functionality. More sophisticated software platforms, the application of machine learning (ML) and artificial intelligence (AI), especially for self-driving capabilities, and more complex in-vehicle computational boards require increased energy consumption~\cite{kothari2024energy}. 

Extended vehicular applications link the vehicle not only to back-end maintenance but also to a full application stack of services, software updates, and data transfer. The most important of these applications are related to Advanced Driver Assistance Systems (ADAS), with shown benefits on road safety \cite{habib2023exploratory}. Fully autonomous driving, again, can be seen as an extended ADAS system requiring even more accurate object recognition, continuous traffic monitoring, and rigorous high-accuracy locationing. These, in turn, require low network latency and, for safety-critical tasks, a local backup decision-making system if no network is available, thereby increasing the in-vehicular computational demands. Further, modern ML/AI technologies invite novel applications not only in safety but also in comfort, personalisation, vehicle-passenger interaction, and infotainment and entertainment.  With increased data and network load, more energy-efficient and sustainable solutions are needed at every level, from the vehicle to the network and cloud infrastructure \cite{montazerolghaem2023efficient}. 

The majority of existing research has focused on understanding hardware components and overall energy use, especially in electric vehicles, where reduced driving range is considered a nuisance and a loss of customer value. Modern vehicular architectures rely on a network of computers, so-called Electronic Control Units (ECUs), with limited memory and CPU capabilities, underscoring the complexity of answering the question "which operation consumes what amount of energy". Harnessing complexity, again, underscores the need for proper energy management that treats hardware and software components equally. Beyond a sustainability perspective, increasing the capacity of ECUs -- and thus enabling more advanced safety features -- would result in more expensive cars, exacerbating financial and societal inequality in traffic safety and increasing material demands for manufacturing more powerful computing hardware. 

Increased software use in modern vehicles can, however, also provide tools for understanding \textbf{holistic energy consumption}. This new type of energy- and sustainability-aware SDV design should focus not only on the vehicle's range optimisation but also on balancing safety-critical features and the energy demands of low-latency ML/AI operations, introducing new applications for passenger comfort and traffic fluency, and maintaining sustainable energy consumption levels. Optimisation, however, requires accurate power models of how different hardware and software components use the energy.

In this paper, we discuss the sustainability perspectives on software-defined vehicle development and provide a vision for software-driven battery modelling utilised by SDV frameworks. We argue that the key requirement for energy modelling is understanding how different software components (microservices, services, and applications) and hardware components (ECUs and sensors) utilise energy, and how SDV designs should be guided towards a more sustainable direction. 
\section{Background: Software-defined Vehicles and Sustainability}

Sustainability of the computing continuum from client devices, whether they are smartphones, wearable devices, traffic sensors, or software-defined vehicles, through the edge of the network, all the way through base stations to the cloud services in the data centres, has been under scrutiny in recent years \cite{peltonen2025rethinking}. Due to highly siloed software ecosystems, most sustainability and energy-efficiency actions have targeted a single category of software products or a specific layer in the continuum, such as data centres. However, in a large-scale continuum of heterogeneous client devices and the exponential growth in capacity demand on the data centre side, focusing on a single layer or an industry-specific silo has become insufficient without a broader view of the bigger picture. This paradigm shift means that every level and its corresponding component, along with their energy efficiency and sustainability footprint, should be known.

Traditionally, road traffic vehicles, especially personal cars, have not been seen as part of software ecosystems, even though modern cars are increasingly software products. Software-defined vehicles (SDVs) utilise extended computational capabilities to organise various operational functions related to driving and road safety, as well as for comfort, entertainment, and traffic functions such as route optimisation \cite{FEDERATE2024}. It is important to note that a software-defined vehicle, by default, does not specify the powertrain solution; it can be based on any fuel type, electric, or hybrid. A software-defined vehicle simply means that its operational orchestration is based on the software-defined paradigm: its operational functions are managed and updated through software rather than a traditional mechanical approach \cite{jiang2024vehicle}. However, for simplicity, we focus on electric powertrains, i.e. electric vehicles (EVs) in this paper. EVs are generally considered the most energy-efficient powertrain \cite{habib2023exploratory}, thereby making the focus more tangible than for fuel-centred powertrains. In addition, it necessary to mention that sustainability on vehicles and road transportation in general is a larger topic beyond energy consumption: vehicle and ECU building materials (especially rare metals), sources of the electricity used for the powertrain (renewable and non-renewable sources), building and ageing of the EV batteries, and overall lifecycle management of the vehicle through its lifespan, all contribute in the sustainability footprint of the vehicle. However, as the software directly affects energy consumption, this perspective is also taken in this paper. 

Some previous research has considered electric vehicle energy consumption in general, categorising, for example, the energy demands of each component, vehicle dynamics, traffic, and environment-related factors \cite{chen2021review}. As vehicles become more complex and versatile, factors beyond the powertrain, such as AC and heating, increasingly influence energy consumption \cite{wager2016driving}. However, software components and their energy demands remain significantly underrepresented in the literature. The effect of network load on vehicles' increasing energy demand has been noted in previous research \cite{montazerolghaem2023efficient}, but solutions focus on optimising and minimising network load rather than on comprehensive, vehicle-perspective energy balancing. For example, as security becomes increasingly crucial for Internet-accessible vehicles, some effort has focused on evaluating the energy consumption of cryptographic methods in in-vehicle networking \cite{kurunathan2025energy}. 

To evaluate the energy consumption of software-defined vehicles as client devices in the computing continuum, a more software-oriented perspective should be adopted for measuring energy usage. Instead of focusing on individual ECUs or components within a given ECU domain, it is necessary to measure the energy each vehicle functionality actually consumes. This way, the energy can be seen as a resource for software quality, similar to computational or available network capacity, and treated as a part of the larger optimisation environment. Some direction in that has already been taken through the design and optimisation of Advanced Driver Assistance Systems (ADAS) \cite{habib2023exploratory}. ADAS systems can reduce energy and fuel consumption across different powertrains \cite{achariyaviriya2024comparative}; however, continuous imaging and object recognition algorithms require increasing computational resources, thereby directly increasing energy demands. As cars become safer and more intelligent, they are at the same time becoming more energy-hungry. Offloading to the network edge can address some of these resource requirements \cite{kothari2024energy}. However, using edge network assistance requires persistent high-bandwidth network access, as required by safety standards. For critical safety considerations, the highest-risk application should also always operate at the vehicle level \cite{FEDERATE2024}. 

A software-oriented perspective on energy usage also enables greater human awareness of the vehicle's energy consumption through range optimisation, as drivers tend to choose routes based on available range. Energy-aware recommendations could then be made to improve the effectiveness of route recommendations, accounting for the different energy impacts of driving on urban roads at lower speeds, well-maintained motorways, or possibly hazardous remote rural roads \cite{wu2015electric}. Such environmental factors can be further expanded to consider the vehicle's "personalised" energy profile, combining driving behaviour information with factors such as vehicle mass, speed, and road gradient \cite{zhang2025extended}. Overall, transparency in energy consumption can improve route choices, either by enabling autonomous vehicles to make better route decisions or by increasing human awareness through proposed route recommendations. For this, a reliable method for creating measurable energy estimates across varying situations needs to be developed. 

On the manufacturer side, circuit design can help produce more energy-efficient sensors and ECU solutions \cite{ragonese2022cmos}. However, from a software development perspective, more understanding is needed on how a given piece of software (service, microservice, application) utilises energy in practical situations. Especially important this is when implementing in-vehicular ML/AI and inference applications that require extended data and computational capacity. By using measurable performance indicators for software development pipelines, a general understanding of vehicular software energy demands can be made more transparent and easier to access for third-party application developers \cite{peltonen2025rethinking}, even further removed from manufacturers (OEMs) and chip providers (Tier 1) in the SDV value chains.

\section{Measuring ADAS Energy Usage: a Preliminary Study}

The powertrain that drives the motor and, quite literally, moves the vehicle is, by all means, the most energy-hungry component in any device moving at motorway speeds. Thus, the underlying question remains: how much energy do the other components, such as driving support systems and air conditioning, actually consume? There are no comprehensive answers to these questions in the research literature, and the manufacturer's estimates may be more positive than what actually happens in real-world scenarios. Most sensor manufacturers' Software Development Kits (SDKs) do not provide a direct interface for energy usage readings. Thus, for experimentation, energy measurements must be performed first manually and then implemented at the operating system (or hypervisor) level within the software architecture. In this section, we present preliminary energy consumption measurements from a simple experimental setup suitable for ADAS and similar applications that utilise remote sensing capabilities. We also review manufacturer and third-party reports to compare our results, establish a benchmark for the need for energy considerations (why this should be done), and later propose a software-oriented pipeline for implementing it in future work. 

\begin{table}[]
    \centering
    \begin{tabular}{p{7cm} | c c}
        \textbf{Device} & \textbf{Idle Watts} & \textbf{40\% load Watts} \\ \hline
        Full setup & 84.6 & 108.1 \\
        Computing: Jetson AGX Orin & 30.55 & 35.25 \\
        Lidar: Hesai OT128 & 35.25 & 42.3 \\
        Stereocamera: Carnegie Robotics Multisense S27 & 16.45 & 23.5 \\
        Thermal cameras (2pcs): Teledyne FLIR ADK 2.0 & 4.7 & 11.75 \\
    \end{tabular}
    \caption{Watts measured on a simple ADAS setup on lab conditions and direct current (235VAC). Load is measured on a computer (Jetson AGX Orin).}
    \label{tab:measurements}
\end{table}

\subsection{Experimental Setup}
Our experimental setup consists of a vehicle-scale computer, lidar, stereocamera, and two thermal cameras. Computational capacity is provided by the NVIDIA Jetson AGX Orin. A 2TB M.2 SSD is attached to it for storage. It has a 12-core ARM Cortex processor, a 2048-core NVIDIA Ampere GPU, and 64GB of internal memory. It offers a range of connectivity options, including an RJ45 connection supporting up to 10GbE, a 40-pin header, and USB connectors. The sensors, including the LiDAR, stereocamera, and two thermal cameras, were interfaced with the Robotic Operating System (ROS2). 

The lidar considered in this study is Hesai OT128, an automotive-grade 360° long-range LiDAR with a 200m range at 10\% reflectivity and a maximum range of 240m. The field of view (FoV) is 360° horizontally and 40° vertically. The stereocamera is the Carnegie Robotics Multisense S27. It features two monochrome sensors at 1920 x 1280 resolution with a global shutter, as well as an RGB sensor for colour capture at 1920 x 1188 resolution with a rolling shutter. Finally, two thermal cameras are Teledyne FLIR ADK 2.0, with a horizontal FoV of 24° and USB-A connections. They have a resolution of 640 x 512, frame rates of 9, 30, or 60 Hz, and can provide either 16-bit TIFF images or compressed 8-bit images.

In our usual research pipelines, the lidar setup is mounted on a real car, but for energy consumption measurements, it was tested in a lab. In a car, electricity would be provided by an inverter (12 VDC and 240 VAC) and a high-capacity battery. In the lab, the electricity was supplied by a standard alternating current (measured at 235 VAC; the standard household wall voltage at the location is 230 VAC). The amperes (A) were measured separately with a Fluke 43 Power Quality Analyser: the full setup, the Jetson computer only, and each sensor on/off. The measurement was repeated with the Jetson Orin idle (only powered on) and with a 40\% CPU load on the Jetson. This load corresponds to the sensor data being read and saved to an external SSD without concurrent processing tasks, such as ML/AI training or inference. The results of the measurements, transferred to watts (W), are reported in Table \ref{tab:measurements}. 

\subsection{Comparison to Existing Knowledge}

There is a significant lack of academic papers that measure vehicle energy consumption from a software perspective; however, some comparisons can be made. Miri et al. \cite{miri2021electric} report electric vehicle (EV) energy costs by operation, including driving control (150W), audio (35W), and power steering (400W). These numbers include the mechanical actuators' functional energy consumption, and it is hard to separate the software-specific energy demands. The NVIDIA Jetson AGX Orin is configurable with power consumption ranging from 15W to 75W, which corresponds to the measurements from our experiment with no energy-saving modes enabled (see Table \ref{tab:measurements}). 

However, non-academic sources estimate that Nvidia DRIVE AGX Orin, a software update provided by Nvidia for their flagship next-generation automotive platform, has reported to come with power consumption varying between 130W and 750W \footnote{\url{https://anyconnect.com/recommended-sbcs/Nvidia/DRIVE-AGX-Orin-Low-Power}}. Similarly, Tesla FSD models HW3 and HW4 have reported 70–150W energy consumption in normal driving, but some automated "robotaxi" level features are estimated to consume 800-1000W on computing capacity alone \footnote{\url{https://www.notateslaapp.com/news/2081/tesla-officially-announces-fsd-hardware-50-and-how-it-compares-to-hardware-40}}. With these estimates, it is critical to note that the estimates are given by either the company representatives or professionals following the EV development field, not by academic independent studies. 

If combined, the comparable energy consumption for an electric vehicle (around 15–20kW per 100 kilometres) and the extra 100W used for simple sensor readings (as presented in Table~\ref{tab:measurements}) would be less than a percentage of the full energy consumption. However, with extended self-driving features and the computational capacity required for ML/AI inference, a 1000W increase in energy demand already represents an additional 5\% increase in energy consumption. As these vehicles also utilise various energy-saving technologies, including recovery propulsion systems \cite{hosseini2023energy}, estimating total energy consumption can be difficult. In addition, environmental factors, such as differences in traffic speed and driving behaviour, affect the total energy consumption at both the hardware and software levels. This motivates further studies, especially those based on real-world data collection and statistical analysis, to better understand the effects of increased vehicle software capabilities. With the development of more intelligent vehicles and applications, the need for additional sensors and computational capacity is increasing, leading to ever-increasing energy consumption \cite{Boulay2020ADAS}.

It is noteworthy that the numbers presented in this section relate only to driving assistance, ADAS, and self-driving features. With extended software capabilities, the vehicles will implement other applications utilising computational capacity and ML/AI inference. These include driver adjustments and personalisation, route optimisation based on personal and other preferences, media and entertainment preferences, and continuous feature delivery with over-the-air updates (OTA) \cite{khamis2025rethinking,montazerolghaem2023efficient}. Any real-time ML/AI operation or inference task incurs energy costs \cite{cavus2025next}: data gathering via sensors (as shown in Table \ref{tab:measurements}), local data processing, and networking costs when tasks are offloaded.

\textbf{Threats to validity:} It is noteworthy that our results on energy measurements (see Table \ref{tab:measurements}) are very preliminary and contain only a limited number of sensors in a limited use case scenario. For example, measuring actual ADAS behaviour would require more work on the experimental setup, including running the test in a car rather than in a lab to enable sensor data to flow. For example, the LiDAR seems to use substantial computational resources, depending on how many measurement points are needed to generate a point cloud. By presenting the preliminary results, we aim to open a discussion on the effects of software components' energy usage on SDV development and to apply the further considerations in our future work. 

\section{Sustainability as a Design Consideration}

As a long-term goal, the vehicular computing tasks should be able to be measured in terms of their resource use so that the sustainability indicators (energy consumption and beyond) would be tangible for developers to understand and transparent to the larger-scale client-edge-cloud continuum, for understanding the systematic level of sustainability performance \cite{peltonen2025rethinking}. In this section, we highlight more detailed design goals for vehicular sustainability, without forgetting the larger-scale level, and view vehicles as the newest addition to the family of Internet-based software applications. 

\textbf{Efficiency vs. added resource use.} The core trade-off to discuss lies in the efficiency of the sensing task versus the additional resources required to achieve the level and quality of performance needed for the task in question. Adding more high-quality sensors that provide more accurate data may save human lives in security- and safety-critical scenarios, such as crash and accident prevention in traffic. However, adding more high-quality sensors that provide more accurate data also increases computing requirements and, inevitably, energy consumption and the amount of required materials. In ethical terms, it can be questioned whether added complexity indeed yields positive net effects, and how sure we can be about the system-level gains. Adding materials and more expensive technology to vehicles reduces traffic congestion and decreases the number of accidents; however, the direct costs (extra hardware, energy use, mining materials) are immediate and concrete. 

\textbf{Societal vs. environmental trade-offs.} Saving lives now in the traffic accidents may, with added energy and material demands, indeed come with long-term environmental harm, which humans are not immune to. The costs of added sensing capabilities in vehicles lie in resource extraction (lithium, rare earths, semiconductors), manufacturing emissions (often in other regions than those where the target consumers live), and e-waste from rapidly obsolete electronics. The same analogy applies to offloading energy-intensive tasks from client devices to the edge of the network or to a distant cloud: cloud environments and data centres are indeed needed for training the most data-intensive AI models, which, in turn, raises several environmental problems. This underscores the need for transparency of energy and sustainability indicators across the entire computing continuum. 

\textbf{Distribution of benefits and harms.} The SDV development for now is first and foremost focused on high-end vehicles that already come with more luxury add-ons. If high-end sensors, such as LiDARs, become more affordable in the future, safety-critical SDV-enabled features, such as extended ADAS, may be more widely adopted in the vehicle pool available to customers of lower economic status. Before then, not everyone is equally affected by the benefits of safer driving: people who can afford newer, more expensive vehicles benefit first. As it is somewhat unrealistic to expect many required rare materials to become significantly more affordable anytime soon, the real societal design question is how to design effective safety solutions with lower sensing capability, such as utilising regular cameras or lower-end, more affordable LiDARs. For analysing the effectiveness of such solutions, more transparent indicators are again needed: in addition to model performance, other factors, such as energy consumption and the pure price of the solutions, should be considered. 

Based on these considerations, the following principles can be delivered as design considerations: 
\begin{enumerate}
    \item The manufacturer's (OEM) responsibility is to aim for sustainable, societally responsible designs that balance energy and resource efficiency, required sensor and material usage, and societal accessibility across different vehicle price ranges. 
    \item Chip provider (Tier 1) responsibility is to aim for energy-efficient designs and awareness in material usage, and balance innovation effects between low and high-end solutions.
    \item Software industry responsibilities aim to design solutions that capture the holistic picture of vehicular computing stacks and their sustainability indicators, avoid redundant technical solutions, and make sustainability indicators transparent to the industry. 
    \item Researchers' and R\&D practitioners' responsibilities aim to define key sustainability indicators that are tangible for the industry to follow, transparent by design, and measurable in ways that enable use within complex, heterogeneous systems, such as vehicular computing architectures. 
\end{enumerate}

% \section{Towards the Measurable Energy Indicators with Microservice-based Approach}

% Suddenly, the SDV paradigm itself proposes the solution to the problem it has created -> Dockers to measure energy

%In the long run, focus has been on simulations and battery emulations due to the high costs of real-world studies \cite{zhang2015electric,miri2021electric}. Installation requires a specific modified vehicle, even if energy models can be built on top of it with some accuracy \cite{wu2015electric}. The model might be vehicle and/or model-specific. 
\section{Discussion}

To fully comprehend the sustainability of software-defined vehicles, novel energy modelling and measurement practices are needed to understand both software and hardware components, and their combined energy consumption, in a truly transparent manner. Even if more affordable and energy-efficient chips are designed and full lifecycle management, including the recycling of rare materials, is implemented, the role of software components also needs to be understood. This means that software architectures need to adapt to implement suitable tools for measuring energy consumption at runtime and other sustainability-related indicators. 

In the current research, most energy modelling is performed using various simulations and battery emulators \cite{zhang2015electric,miri2021electric}. This is often due to the high costs of real-world studies, including suitable test vehicles and the infrastructure needed to run them, such as test tracks and expertise in building data-collection interfaces. In most cases, data collection from an actual vehicle requires installing equipment or a specifically modified vehicle that has a physically opened reader for a vehicular networking environment (e.g., CANbus access). In the end, the model might still be vehicle and/or model-specific and hard to generalise over multiple manufacturers, models, or powertrains \cite{wu2015electric,zhang2015electric}. The simulations overcome these real-world challenges but leave open the question of whether the models actually capture the versatility of real-life driving scenarios, the heterogeneity of SDV architectures, and the effects of environmental factors in the long run \cite{zhang2025extended}. Thus, we recommend that the solution utilise the best practices from both worlds: simulations and emulations, extended with real-world data use cases and novel measurement technologies that integrate crowdsensing and crowdsourcing into the vehicular domain. This would, however, require fully operational SDV fleets already in use and suitable software security and privacy measures in place to address data privacy challenges. 

\begin{figure}
    \centering
    \includegraphics[width=\linewidth]{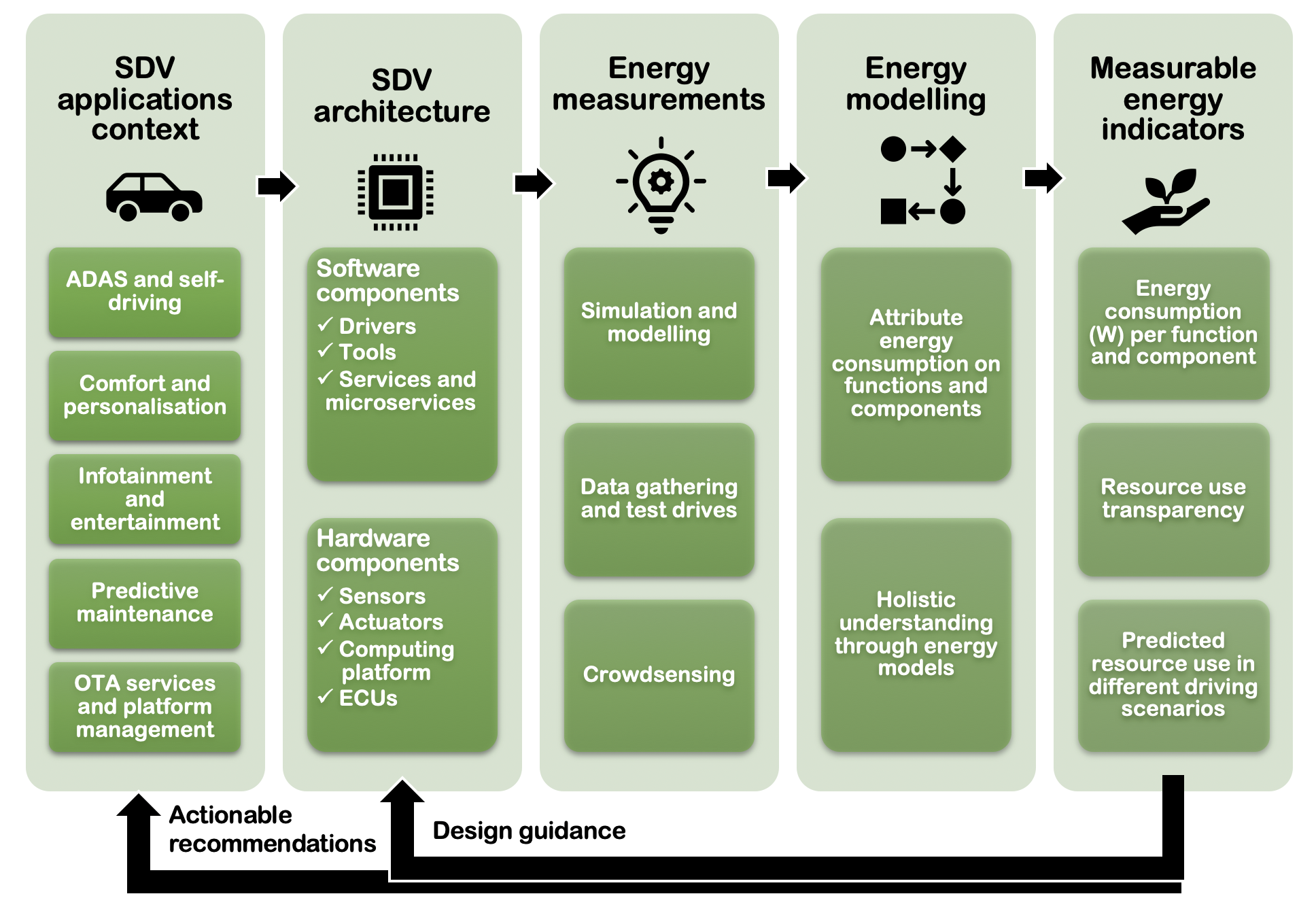}
    \caption{An overview of the SDV applications and architectures and how they would benefit from energy measurements, modelling, and sustainability indicators through actionable energy-saving recommendations and sustainable design guidance. }
    \label{fig:overview}
\end{figure}

Picture \ref{fig:overview} provides an overview of SDV applications and architectures and their relation to energy measurement, modelling, and the definition of measurable sustainability indicators. The overview can also be considered a research agenda: given the application context, architectures should be designed to enable resource usage (including energy and other sustainability indicators) to be measured at the service level. This can be realised, for example, through microservice architectures that enable energy management, measurement, and optimisation tools \cite{araujo2024energy}, which are already widely adopted across multiple cyber-physical systems \cite{mena2023towards}, including vehicles. Indeed, the industrial development of SDVs is already moving towards service-oriented architectures \cite{FEDERATE2024}. However, implementation and orchestration, including the delivery of tangible SDKs that integrate energy-measurement features, are yet to be completed.

For the evaluation and validation of energy measurements, more research is required, utilising mixed methods that range from simulations and emulations to real-world data studies measuring actual vehicles in real consumer environments. With enough data, energy usage modelling can be implemented to create measurable energy indicators that shall provide both design guidance for software developers and actionable recommendations for end-users. Through the modelling and recommendation feedback loop, both new innovations in energy management and more energy-efficient SDV architectures and services should emerge. 

\section{Conclusions}

In this paper, we have covered the sustainability challenges in software-defined vehicle development. As mechanical entities such as cars become part of the Internet-driven value chains and architectures, the so-called computing continuum, more focus should be given to how novel software solutions and hardware-level chip designs can better integrate sustainability and energy management principles. In addition to viewing the vehicle as a client device, it is crucial to understand its role within the continuum and to provide transparent, measurable indicators for assessing energy efficiency and other sustainability goals. In this paper, we have presented a preliminary independent study to highlight that software-driven energy usage indeed remains an unsolved resource optimisation problem that can influence the vehicle's overall energy consumption. This highlights the underlying ethical considerations in SDV development, where integrating more sophisticated technologies to enhance human safety can exacerbate broader environmental and societal problems. On the path to making modern vehicles safer and smarter, it is crucial that they not become more energy- and resource-hungry at the same time.

\begin{acknowledgments}
The work has been supported by the EU HORIZON projects CHIPS-JU CIA FEDERATE (grant number 101139749), CHIPS-JU RIA HAL4SDV (grant number 101139789) and CHIPS-JU IA SHIFT2SDV (grant number 101194245), and Business Finland national funding for HAL4SDV (grant number 7655/31/2023) and SHIFT2SDV (5972/31/2024). 

Co-funded by the European Union. Views and opinions expressed are, however, those of the author(s) only and do not necessarily reflect those of the European Union or the Chips Joint Undertaking. Neither the European Union nor the granting authorities can be held responsible for them. 
\end{acknowledgments}

%% The declaration on generative AI comes in effect
%% https://ceur-ws.org/GenAI/Policy.html
\section*{Declaration on Generative AI}
 Grammarly was used for grammar and spelling checks. All content was reviewed and finalised by the author(s).
  \newline
%%
%% Define the bibliography file to be used
\bibliography{refs}

@article{peltonen2025rethinking,
  title={Rethinking computing systems in the era of climate crisis: A call for a sustainable computing continuum},
  author={Peltonen, Ella and Bayhan, Suzan and Bermbach, David and Buschjager, Sebastian and Degeler, Victoria and Ding, Aaron Yi and Incel, Ozlem Durmaz and Katare, Dewant and Kjargaard, Mikkel Baun and Leroux, Sam and others},
  journal={IEEE Internet Computing},
  year={2025},
  publisher={IEEE}
}

@article{wu2015electric,
  title={Electric vehicles’ energy consumption measurement and estimation},
  author={Wu, Xinkai and Freese, David and Cabrera, Alfredo and Kitch, William A},
  journal={Transportation Research Part D: Transport and Environment},
  volume={34},
  pages={52--67},
  year={2015},
  publisher={Elsevier}
}

@article{chen2021review,
  title={A review and outlook on energy consumption estimation models for electric vehicles},
  author={Chen, Yuche and Wu, Guoyuan and Sun, Ruixiao and Dubey, Abhishek and Laszka, Aron and Pugliese, Philip},
  journal={SAE International Journal of Sustainable Transportation, Energy, Environment, \& Policy},
  volume={2},
  number={13-02-01-0005},
  pages={79--96},
  year={2021}
}

@article{wager2016driving,
  title={Driving electric vehicles at highway speeds: The effect of higher driving speeds on energy consumption and driving range for electric vehicles in {Australia}},
  author={Wager, Guido and Whale, Jonathan and Braunl, Thomas},
  journal={Renewable and sustainable energy reviews},
  volume={63},
  pages={158--165},
  year={2016},
  publisher={Elsevier}
}

@article{zhang2015electric,
  title={Electric vehicles’ energy consumption estimation with real driving condition data},
  author={Zhang, Rui and Yao, Enjian},
  journal={Transportation Research Part D: Transport and Environment},
  volume={41},
  pages={177--187},
  year={2015},
  publisher={Elsevier}
}

@article{miri2021electric,
  title={Electric vehicle energy consumption modelling and estimation—A case study},
  author={Miri, Ily{\`e}s and Fotouhi, Abbas and Ewin, Nathan},
  journal={International Journal of Energy Research},
  volume={45},
  number={1},
  pages={501--520},
  year={2021},
  publisher={Wiley Online Library}
}

@article{montazerolghaem2023efficient,
  title={Efficient resource allocation for multimedia streaming in software-defined internet of vehicles},
  author={Montazerolghaem, Ahmadreza},
  journal={IEEE Transactions on Intelligent Transportation Systems},
  volume={24},
  number={12},
  pages={14718--14731},
  year={2023},
  publisher={IEEE}
}

@inproceedings{zhang2025extended,
  title={{Extended vehicle energy dataset (eVED): An enhanced large-scale dataset for vehicle energy consumption analysis}},
  author={Zhang, Shiliang and Fatih, Dyako and Abdulqadir, Fahmi and Schwarz, Tobias and Ma, Xuehui},
  booktitle={2025 IEEE 101st Vehicular Technology Conference (VTC2025-Spring)},
  pages={01--07},
  year={2025},
  organization={IEEE}
}

@inproceedings{kurunathan2025energy,
  title={{Energy Profiling of Lightweight Authentication protocols for Software-Defined Vehicles}},
  author={Kurunathan, Harrison and Ali, Hazem Ismail and Eldefrawy, Mohamed Hamdy and Tovar, Eduardo},
  booktitle={2025 12th International Conference on Future Internet of Things and Cloud (FiCloud)},
  pages={537--542},
  year={2025},
  organization={IEEE}
}

@inproceedings{kothari2024energy,
  title={Energy-efficient and context-aware computing in software-defined vehicles for advanced driver assistance systems (ADAS)},
  author={Kothari, Aadi and Talty, Timothy and Huxtable, Scott and Zeng, Haibo},
  booktitle={WCX SAE World Congress Experience},
  year={2024},
  organization={SAE Technical Paper}
}

@inproceedings{habib2023exploratory,
  title={{An Exploratory Research on Electric Vehicle Sustainability: An Approach of ADAS}},
  author={Habib, M and Mithu, A and Zihad, F},
  booktitle={1st International Conference on Smart Mobility and Vehicle Electrification},
  year={2023}
}

@inproceedings{achariyaviriya2024comparative,
  title={Comparative analysis of energy consumption in semi-autonomous vehicles: The influence of adaptive cruise control across various powertrains},
  author={Achariyaviriya, Witsarut and Janpoom, Kittitat and Wanison, Ramnarong and Mona, Yuttana and Wongsapai, Wongkot and Kammuang-Lue, Niti and Suttakul, Pana},
  booktitle={AIP Conference Proceedings},
  volume={3236},
  number={1},
  pages={080004},
  year={2024},
  organization={AIP Publishing LLC}
}

@article{Boulay2020ADAS,
  author    = {Boulay, Pierre and Malaquin, Christophe},
  title     = {{ADAS Sensors and Computing: A 22 Billion Dollar Market in 2025}},
  journal   = {PhotonicsViews},
  year      = {2020},
  volume    = {17},
  number    = {4},
  pages     = {35--37},
  doi       = {10.1002/phvs.202070410},
  publisher = {Wiley-VCH}
}

@article{ragonese2022cmos,
  title={{CMOS automotive radar sensors: mm-Wave circuit design challenges}},
  author={Ragonese, Egidio and Papotto, Giuseppe and Nocera, Claudio and Cavarra, Andrea and Palmisano, Giuseppe},
  journal={IEEE Transactions on Circuits and Systems II: Express Briefs},
  volume={69},
  number={6},
  pages={2610--2616},
  year={2022},
  publisher={IEEE}
}

@article{hosseini2023energy,
  title={Energy recovery and energy harvesting in electric and fuel cell vehicles, a review of recent advances},
  author={Hosseini, Seyed Mohammad and Soleymani, Mehdi and Kelouwani, Sousso and Amamou, Ali Akrem},
  journal={IEEE Access},
  volume={11},
  pages={83107--83135},
  year={2023},
  publisher={IEEE}
}

@article{khamis2025rethinking,
  title={Rethinking Vehicle Architecture Through Softwarization and Servitization},
  author={Khamis, Alaa and Goswami, Partha},
  journal={IEEE Access},
  volume={13},
  pages={126213--126226},
  year={2025}
}

@article{cavus2025next,
  title={Next generation of electric vehicles: AI-driven approaches for predictive maintenance and battery management},
  author={Cavus, Muhammed and Dissanayake, Dilum and Bell, Margaret},
  journal={Energies},
  volume={18},
  number={5},
  pages={1041},
  year={2025},
  publisher={MDPI}
}

@inproceedings{jiang2024vehicle,
  title={Vehicle e/e architecture and key technologies enabling software-defined vehicle},
  author={Jiang, Shugang},
  booktitle={WCX SAE World Congress Experience},
  year={2024},
  organization={SAE Technical Paper}
}

@misc{FEDERATE2024,
  author       = {{FEDERATE Consortium and SDVoF Sherpa Governance Team}},
  title        = {{European Software-Defined Vehicle of the Future (SDVoF) Initiative – Vision and Roadmap}},
  year         = 2024,
  month        = apr,
  version      = {17},
url          = {federate-sdv.eu},
  @url          = {federate-sdv.eu/2024/04/12/european-software-defined-vehicle-of-the-future-sdvof-initiative-vision-and-roadmap/},
  note         = {Accessed: 2026-05-14}
}

@article{araujo2024energy,
  title={Energy consumption in microservices architectures: a systematic literature review},
  author={Ara{\'u}jo, Gabriel and Barbosa, Vandirleya and Lima, Luiz Nelson and Sabino, Arthur and Brito, Carlos and F{\'e}, Iure and Rego, Paulo and Choi, Eunmi and Min, Dugki and Nguyen, Tuan Anh and others},
  journal={IEEE Access},
  volume={12},
  pages={186710--186729},
  year={2024},
  publisher={IEEE}
}

@article{mena2023towards,
  title={Towards high-availability cyber-physical systems using a microservice architecture},
  author={Mena, Manel and Criado, Javier and Iribarne, Luis and Corral, Antonio and Chbeir, Richard and Manolopoulos, Yannis},
  journal={Computing},
  volume={105},
  number={8},
  pages={1745--1768},
  year={2023},
  publisher={Springer}
}

\end{document}